\documentclass{camera}
\usepackage{epsfig}

\begin{document}

%%%%%%%%%%%%%%%%%%%%%%%%%%%%%%%%%%%%%%%%%%%%%%%%%%%%%%%%
% The title, all uppercase; if you want to split it in
% two or more lines, put a \\ macro at each line break
% example: 
%   \title{TITLE: FIRST LINE\\ SECOND LINE}
%
\title{RITORNO RADIATIVO PER LA MISURA DI R: 
COME E PERCH\'E~\footnote{Talk given in italian at Incontri sulla 
Fisica delle Alte Energie, Lecce, Italy, Apr 2003.}}

%%%%%%%%%%%%%%%%%%%%%%%%%%%%%%%%%%%%%%%%%%%%%%%%%%%%%%%%
% The author(s), separated by commas; do not put a
% comma before the last author, use instead the \and
% macro which produces a normal ``and'' in the
% caps/small caps context
%
\author{Germ\'an Rodrigo~\footnote{Supported by EC 5th Framework Programme 
under contract HPMF-CT-2000-00989.
E-mail: {\tt german.rodrigo@cern.ch}}}

%%%%%%%%%%%%%%%%%%%%%%%%%%%%%%%%%%%%%%%%%%%%%%%%%%%%%%%%
%
\organization{Theory Division, CERN, CH-1211 Geneva 23, Switzerland}

\maketitle

\vspace{-8cm}
\hfill {\tt CERN-TH/2003-275}

\vspace{7.5cm}

%%%%%%%%%%%%%%%%%%%%%%%%%%%%%%%%%%%%%%%%%%%%%%%%%%%%%%%%
% Write the text starting from here and using the usual
% LaTeX commands.
%
\begin{abstract}
The measurement of the pion form factor and, more
generally, of the cross section for electron--positron annihilation
into hadrons through the radiative return has become an important
task for high luminosity colliders such as the $\Phi$- or
$B$-meson factories. This quantity is crucial for predictions of 
the hadronic contributions to $(g-2)_\mu$, the anomalous magnetic 
moment of the muon, and to the running of the electromagnetic coupling.
But the radiative return opens also the possibility of many 
other physical applications. The physics potential of this method at 
high luminosity meson factories is discussed and recent results
are reviewed.
\end{abstract}

Electron--positron annihilation into hadrons is one of the basic
reactions of particle physics, crucial for the understanding
of hadronic interactions. At high energies, around the $Z$ resonance,
the measurement of the inclusive cross section and its interpretation
within perturbative QCD give rise to one of the most
precise and theoretically founded  determinations of the strong coupling
constant $\alpha_s$. Also, measurements in the intermediate 
energy region, between 3 GeV and 11 GeV can be used to determine $\alpha_s$
and at the same time give rise to precise measurements of charm 
and bottom quark masses. The low energy region is crucial
for predictions of the hadronic contributions to $a_\mu=(g-2)_\mu/2$, 
the anomalous magnetic moment of the muon, and to the running of the electromagnetic
coupling from its value at low energy up to $M_Z$. 
Last, but not least, the investigation of the exclusive final states
at large momenta allows for tests of our theoretical 
understanding of form factors within the framework of perturbative QCD.
Beyond the intrinsic interest in this reaction,
these studies may provide important clues for the interpretation
of exclusive decays of B-mesons, a topic of evident importance for
the extraction of CKM matrix elements. 

The main uncertainty to $a_\mu$ and $\alpha_\mathrm{QED}$ is driven 
by their respective hadronic contributions, which are not calculable
perturbatively, but can be estimated though dispersion integrals 
\begin{equation} 
a_{\mu}^{\mathtt{had},\mathrm{LO}}
= \left( \frac{\alpha m_\mu}{3 \pi} \right)^2 \int_{4m_{\pi}^2}^{\infty}  
\frac{ds}{s^2} \; \hat{K}(s) \; R(s)~, \nonumber
\end{equation}
\begin{equation}
\Delta \alpha_{\mathtt{had}}(m_Z^2) = -\frac{\alpha m_Z^2}{3 \pi} \ 
\mathrm{Re} \int_{4m_{\pi}^2}^{\infty}  
\frac{ds}{s} \; \frac{ R(s)}{s-m_Z^2-i\eta}~, 
\end{equation} 
where the spectral function $R(s)$ is obtained from
experimental data of the reaction $e^+e^- \to$ hadrons.
The most recent experimental result for $a_\mu$~\cite{Bennet}
shows a 2$\sigma$ discrepancy with respect to the SM prediction 
for this quantity~\cite{Davier:2002dy,HMNT02,Ghozzi:2003yn}. 
Alternatively, one can also use current conservation (CVC)
and isospin symmetry to obtain $R(s)$ from $\tau$ decays.   
In the latter, a 0.7$\sigma$ discrepancy is found~\cite{Davier:2002dy},
which however is incompatible with the $e^+e^-$ based result.
Unaccounted isospin breaking corrections due to the difference of the 
mass and width of the neutral to the charged $\rho$-meson
could explain this discrepancy~\cite{Ghozzi:2003yn}, leaving 
the $e^+e^-$ based analysis as the most reliable. 
 
The recent advent of $\Phi$- and $B$-meson factories
allows us to exploit the radiative return to explore the hadronic 
cross section in the whole energy region from threshold up to the nominal energy 
of the collider in one homogeneous data sample~\cite{Binner:1999bt,Zerwas}. 
The radiative suppression factor ${\cal O}(\alpha/\pi)$ is easily 
compensated at these factories by their enormous luminosity. 

\begin{figure}[htb]
\begin{center}
\epsfig{file=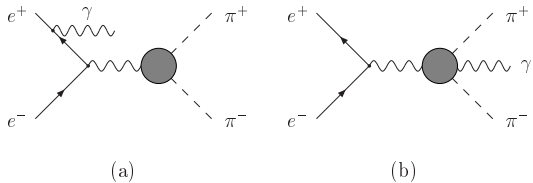,width=7.5cm} 
\caption{Leading order contributions to the reaction 
$e^+e^-\to\pi^+\pi^-\gamma$ from ISR~(a) and FSR~(b).}
\label{fig1}
\end{center}
\end{figure}

In principle, the reaction $e^+e^- \to \gamma + {\mathrm {hadrons}}$ 
receives contributions from both initial- and final-state radiation 
(Fig.~\ref{fig1}), ISR and FSR respectively. Only the former is of 
interest for the radiative return.
A variety of methods to disentangle FSR from the ISR contribution 
have been described in detail in~\cite{Binner:1999bt,Kuhn:2002xg,Rodrigo:2002hk,Czyz:2002np}, 
among them the employment of suitable kinematical cuts to suppress FSR, or the identification 
of different distributions, e.g. angular distributions, charge asymmetry, 
for independent tests of the FSR model amplitude. Notice however that 
at $B$-factories the $\pi^+\pi^-\gamma$ final state is completely 
dominated by ISR.

The proper analysis requires necessarily the construction of Monte Carlo 
event generators. The event generators EVA~\cite{Binner:1999bt} and 
EVA4$\pi$~\cite{Czyz:2000wh} were based on a leading order treatment of 
ISR and FSR, supplemented by an approximate inclusion of additional 
collinear radiation based on structure functions.
Subsequently, the event generator PHOKHARA was developed; it is based
on a complete next-to-leading order (NLO) treatment of 
radiative corrections~\cite{Kuhn:2002xg,Rodrigo:2002hk,Czyz:2002np}.
In its version 2.0 it included ISR at NLO and FSR at LO
for $\pi^+ \pi^-$ and $\mu^+ \mu^-$ final states, and four-pion
final states (without FSR) with some improvements with respect 
to the formulation described in~\cite{Czyz:2000wh}.

The most recent version of PHOKHARA, version 3.0~\cite{Czyz:2002np}, 
allows for the simultaneous emission of one photon from 
the initial and one photon from the final state. 
This includes in particular the radiative return
to $\pi^+ \pi^- (\gamma)$ and thus the measurement of the (one-photon) 
inclusive $\pi^+ \pi^-$ cross section, an issue closely 
connected to the question of $\pi^+ \pi^- (\gamma)$ 
contributions to $a_{\mu}$.

Recently, a new Monte Carlo event generator, EKHARA~\cite{Czyz:2003gb},  
has been constructed to simulate the reaction $e^+e^- \to \pi^+\pi^-e^+e^-$,
a potential background of the radiative return specially at lower energies.  
Future developments of PHOKHARA include the simulation of FSR at NLO 
and the narrow resonances for $\mu^+\mu^-$, as well as many other 
hadronic channels: $K^+K^-$, $K^0\bar{K}^0$, $3\pi$, $KK\pi$, $p\bar{p}$,
and the simulation of the continuum $q\bar{q}$ supplemented by some 
hadronization model. Encouraging preliminary experimental results 
from KLOE, BABAR and BELLE~\cite{unknown:2003jn,Pisa} demonstrate 
the power of the method and its physics potential.\\

{\bf Acknowledgements:}
This contribution is based on work performed in collaboration 
with H.~Czy\.z, A.~Grzeli\'nska and J.H.~K\"uhn. I'm very grateful
to P.~Gambino for his kind invitation and to P.~Ciafaloni for the 
pleasant organization of the workshop. 
Partial support from Generalitat Valenciana under grant CTIDIB/2002/24, 
and MCyT under grant FPA-2001-3031.

\pagebreak 

%%%%%%%%%%%  BIBLIOGRAFIA  %%%%%%%%%%%%

%%%%%%%%%%%%%%%%%%%%%%%%%%%%%%%%%%%%%%%%%%%%%%%%%%%%%%%%
% End of the paper
%
\end{document}